\documentclass[journal=jacsat,manuscript=article]{achemso}

\usepackage{chemformula} 
\usepackage[T1]{fontenc} 
\usepackage{amsfonts} 
\usepackage{jabbrv}
\DefineJournalAbbreviation{Journal of the American Chemical Society}{J A C S} 

\usepackage{qtree}
\usepackage{textcomp}
\usepackage{gensymb} 

\usepackage[subpreambles]{standalone}
\usepackage{siunitx}

\usepackage{graphicx}
\usepackage{dcolumn}
\usepackage{bm}
 \usepackage{color}
 
\usepackage{diffcoeff}  

\usepackage{gensymb} 

\usepackage{braket}

\DeclareUnicodeCharacter{202F}{\,}
\DeclareUnicodeCharacter{2212}{\,}
\DeclareUnicodeCharacter{0301}{\'{e}}
\DeclareUnicodeCharacter{0308}{\"{o}}

\newcommand{\fai}{\varphi}

\newcommand{\br}{{\bf r}}
\newcommand{\bk}{{\bf k}}

\newcommand{\eg}{\textit{e.g.}}
\newcommand{\ie}{\textit{i.e.}}

\newcommand{\viz}{\textit{viz.}}
\newcommand{\etal}{\textit{et al. }}

\usepackage{soul} 

\usepackage{acronym}
\acrodef{MD}{molecular dynamics}
\acrodef{QM}{quantum-mechanical}
\acrodef{SM}{statistical mechanics}
\acrodef{KS-DFT}{Kohn-Sham density functional theory}
\acrodef{FMT}{fundamental measure theory} 
\acrodef{AIMD}{\emph{ab initio} molecular dynamics}
\acrodef{ML}{machine learning}
\acrodef{cDFT}{classical density functional theory}
\acrodef{OZ}{Ornstein-Zernike} 
\acrodef{IET}{integral-equation theory}
\acrodef{SFT}{statistical field theory}
\acrodef{GP}{Gaussian process}

\usepackage{comment}

\usepackage[scr]{rsfso}
\newcommand{\powerset}{\raisebox{.15\baselineskip}{\Large\ensuremath{\wp}}}

\author{Jianzhong Wu}
\email{jwu@engr.ucr.edu}
\affiliation[UCR]
{Department of Chemical and Environmental Engineering, University of California, Riverside, CA 92521, USA}

\author{Mengyang Gu}%
\email{mengyang@pstat.ucsb.edu}
\affiliation [UCSB]{Department of Statistics and Applied Probability, University of California, Santa Barbara, CA 93106, USA}
\title[An \textsf{achemso} demo]
  {Perfecting Liquid-State Theories with Machine Intelligence} 
  
\abbreviations{QM,SM,DFT, ML}

\begin{document}

\begin{tocentry}
  \includegraphics[scale=0.5]{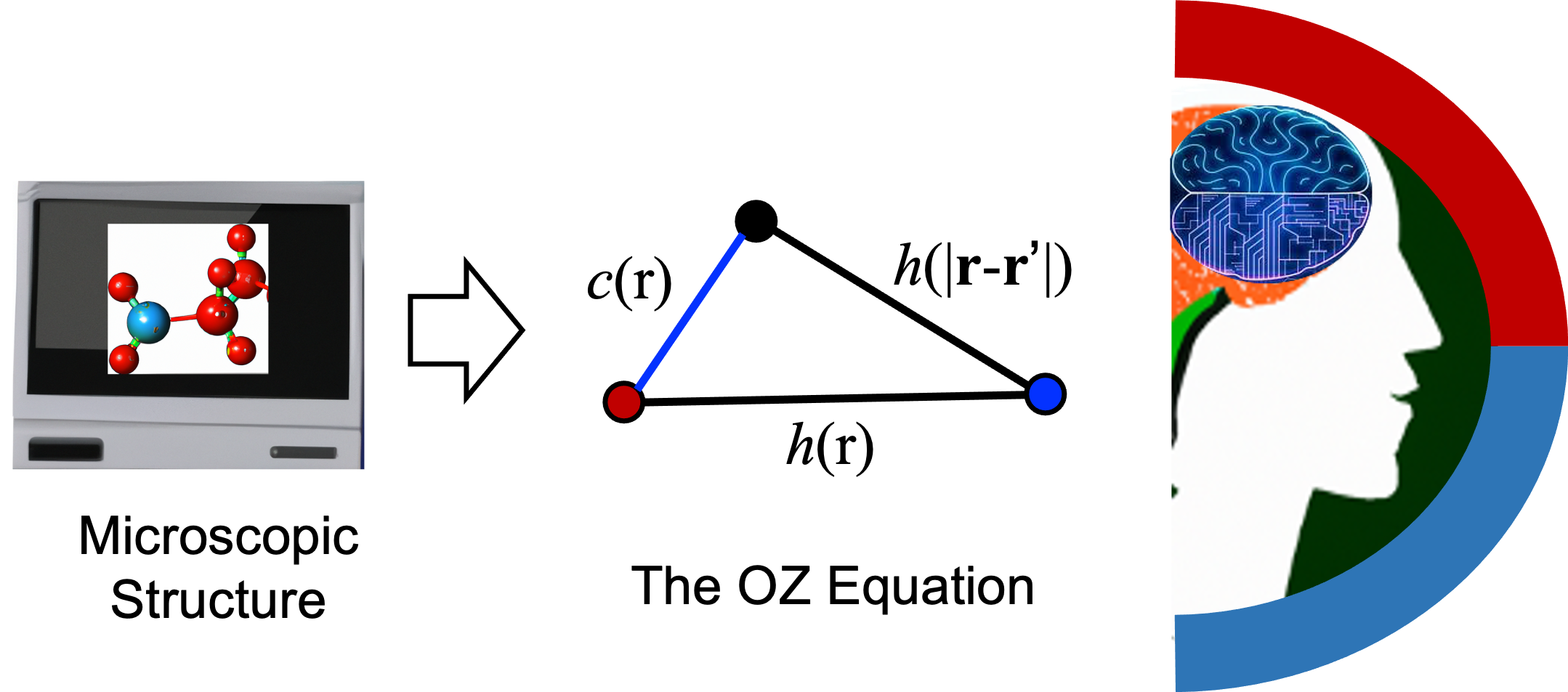}
\end{tocentry}


\begin{abstract}
Recent years have seen a significant increase in the use of machine intelligence for predicting electronic structure, molecular force fields, and the physicochemical properties of various condensed systems. However, substantial challenges remain in developing a comprehensive framework capable of handling a wide range of atomic compositions and thermodynamic conditions. This perspective discusses potential future developments in liquid-state theories leveraging on recent advancements of functional machine learning. By harnessing the strengths of theoretical analysis and machine learning techniques including {{surrogate models}}, dimension reduction and uncertainty quantification, we envision that liquid-state theories will gain significant improvements in accuracy, scalability and computational efficiency, enabling their broader applications across diverse materials and chemical systems. 
\end{abstract}

\section{Introduction}
\begin{figure*}[ht]
    \centering
    \includegraphics[scale=0.6]{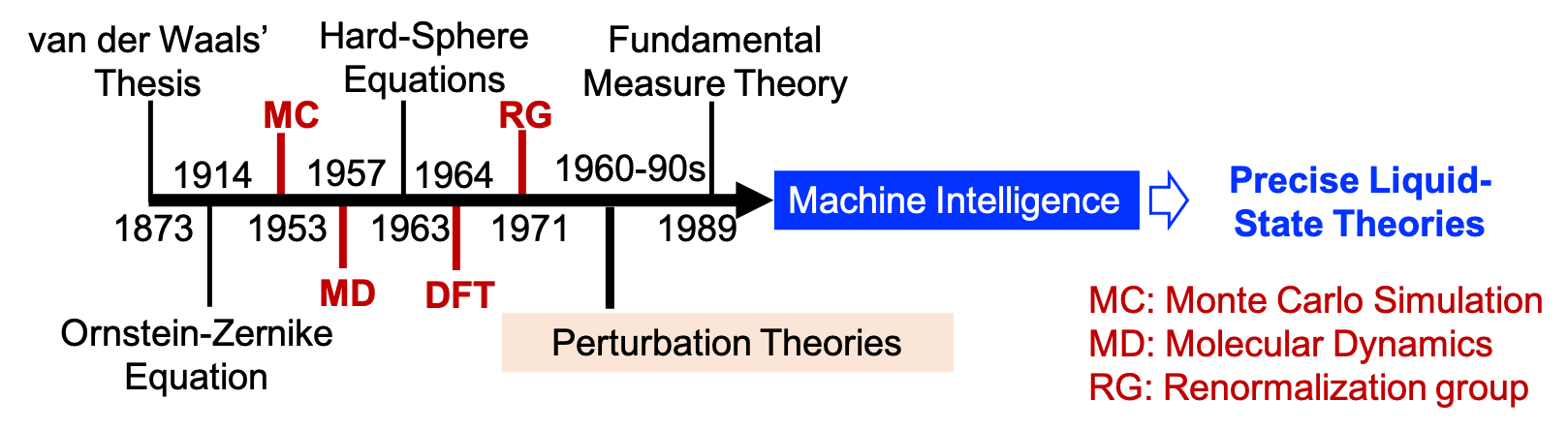}
    \caption{Milestones in the development of liquid-state theories and prospect of future advancement driven by machine intelligence.}
    \label{fig:PerfectTheory}
\end{figure*}

A conventional wisdom in the molecular theory of simple liquids is that the microscopic structure is  determined by molecular excluded volume effects, whereas longer-ranged interactions make contributions to thermodynamic properties that can be adequately described by mean-field approximation.\cite{10.1063/PT.3.4135} This idea may trace its origin back to the pioneering work by Johannes van der Waals on the continuity of the gaseous and liquid states published in 1873. Because the intermolecular repulsion resembles the interaction between billiard balls, the hard-sphere model had been employed to characterize the microscopic structure of simple fluids long before the advent of Monte Carlo (MC) and \ac{MD} simulation methods. Figure \ref{fig:PerfectTheory} illustrates some major milestones in the development of modern liquid-state theories. It is probably not an exaggeration to state that molecular theories of liquid systems are rooted in the mathematical models for predicting the properties of hard spheres.

Hard spheres do not interact with each other unless there is an overlap. As a result, the structure of a hard-sphere system cannot be described with mean-field approximation or typical perturbation methods. While the thermodynamic properties of hard-sphere systems can be obtained through experiment or simulation, analytical theories are mostly derived from geometric analysis, integral-equation theories, and various interpolation schemes.\cite{RN8359} The \ac{FMT}, originally formulated by Yaakov Rosenfeld, provides probably the most systematic and accurate description of the microscopic structure and properties of hard-sphere systems at either uniform or inhomogeneous conditions.\cite{Roth_2010}    

Given an analytical expression for the radial distribution functions of hard-sphere fluids and that for a pairwise additive potential describing intermolecular interactions, the van der Waals picture is capable of predicting the properties of simple liquids in the vicinity of the triple point.\cite{doi:10.1126/science.220.4599.787}  The mean-field approximation can also be used to account for long-range interactions in molecular fluids and electrolyte solutions through a judicious selection of a reference system that reproduces the liquid structure.\cite{doi:10.1021/acs.jpcb.2c06988} While the distinct roles played by the intermolecular short- and long-range interactions in determining the properties of a dense fluid offer valuable insights, van der Waals-type theories encounter challenges when applied to associating fluids and liquid mixtures.\cite{gray2011theory} As a matter of fact, its performance deteriorates even for one-component simple liquids as the thermodynamic condition departs from the triple point. The mean-field approximation ignores fluctuation effects associated with the variation of the local structure due to intermolecular attraction. Such effects can be adequately described with the Barker-Henderson perturbation theory\cite{RN72} or, more systematically, the hierarchical reference theory.\cite{doi:10.1080/00268976.2012.666573} Alternatively, the influence of long-range interactions on the microscopic structure can be explicitly accounted for by employing the functional Taylor expansion of the radial distribution function.\cite{RN2348} The perturbation methods are naturally applicable to one-component fluids as well as mixtures. In combination with the renormalization group (RG) theory to incorporate critical fluctuations,\cite{behnejad2010thermodynamic} they provide quantitative descriptions of thermodynamic properties and phase behavior over the entire range of the vapor-liquid coexistence. Importantly, an explicit consideration of long-range fluctuations is instrumental to understand the asymptotic features of phase transition in pure fluids and mixtures.

In contrast to simple liquids, molecular systems involve intra- and intermolecular interactions as characterized by bond connectivity, electrostatic forces as well as van der Waals interactions. The complexity of the potential energy necessitates the use of more sophisticated theoretical models and advanced computational techniques in order to accurately describe the microscopic structure and thermodynamic properties.\cite{gray2011theory} Such models often rely on semi-empirical force fields or coarse-grained representations of intermolecular interactions. While these classical models play a crucial role in molecular simulation, which yields exact results within the model, practical applications often rely on analytical equations to strike a balance between computational efficiency and accuracy. This balance is essential in various applied fields, such as materials science, drug development, and chemical engineering, where real-world applications demand a pragmatic approach.\cite{doi:10.1021/acs.macromol.6b00107,doi:10.1021/acs.jcim.7b00389,doi:10.1021/acs.iecr.3c02255} 

From a fundamental perspective, one of the most enduring challenges in predicting the properties of liquid systems arises from the need for an accurate description of intermolecular interactions. Although \ac{AIMD} has the potential to predict the multi-body energy for any atomic system, its  utility is severely limited by the computational expense, especially when conducting free-energy calculations as required for constructing the phase diagram of complex molecular systems.  While semi-empirical force fields have been well-established to provide a systematic description of the atomic forces and total energies of molecular systems, their predictive power is constrained by specific functional forms and parameters that must be fitted with experimental data and/or first-principles predictions. Because the classical models frequently invoke the assumption of pairwise additivity for intermolecular interactions, there are also difficulties associated with transferring the model parameters between different molecular systems or thermodynamic conditions. The non-additive nature of many-body interactions implies that the classical models may not be able to provide an accurate description of the molecular energy in both low-density vapor and condensed states such as liquid and solid. Interestingly, the ``short blanket dilemma'', {{\ie, the difficulties for an accurate description of both two-body interactions and thermodynamic properties of condensed phases,}} continues to be a challenge in recent advancements of \ac{ML} force fields.\cite{10.1063/5.0142843} 

\begin{figure*}[ht]
    \centering
    \includegraphics[scale=0.5]{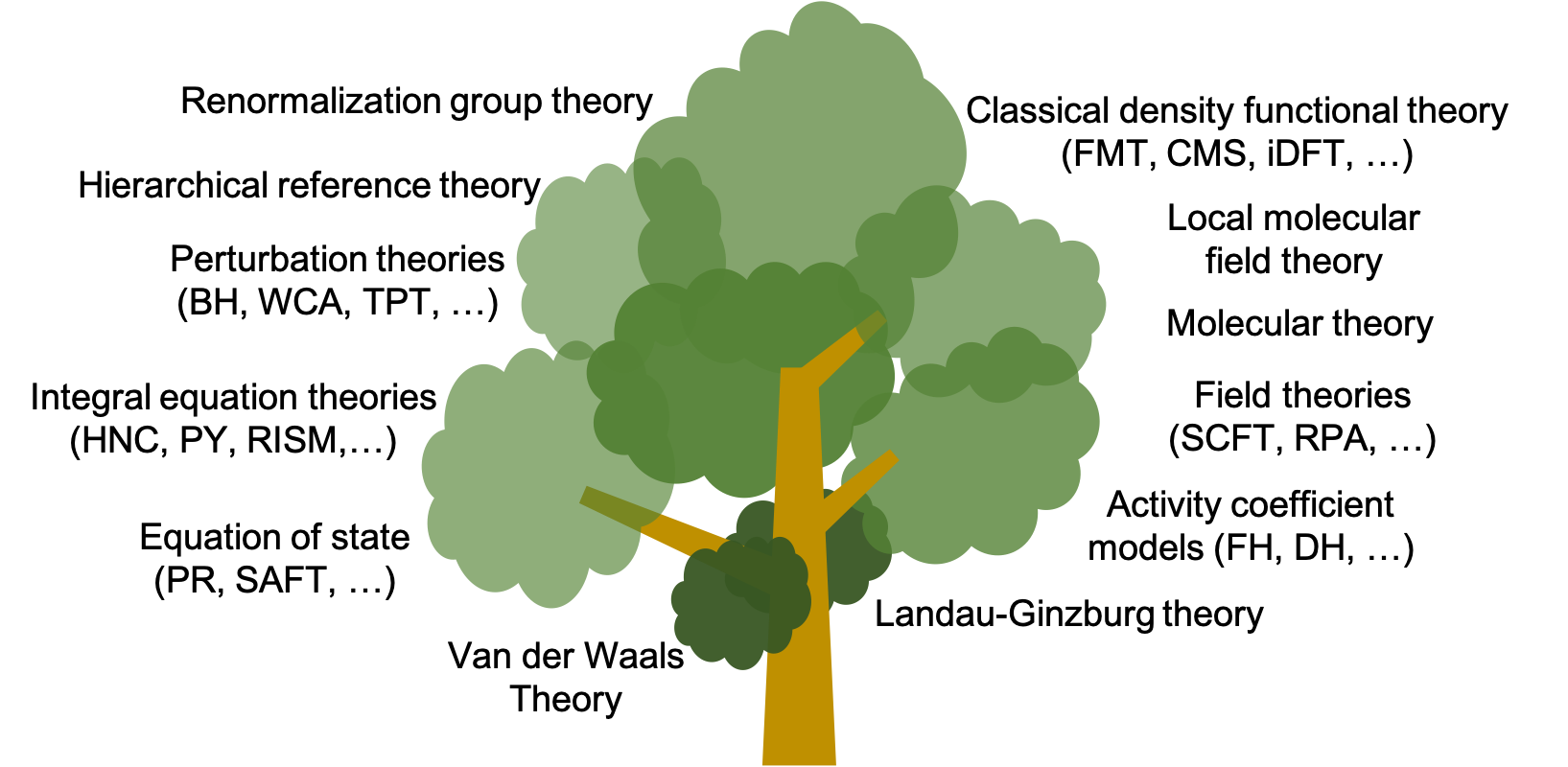}
    \caption{Various forms of liquid-state theories for predicting the thermodynamic properties of molecular systems. The list is incomplete and meant to be illustrative.}
    \label{fig:tree}
\end{figure*}

The potential energy of a molecular system can be highly complex and may not be easily scalable or transferable at the microscopic level. Fortunately, these issues tend to diminish in significance as the system size approaches the thermodynamic limit. In this regard, liquid-state theories offer distinct advantages when compared to molecular simulation because they are naturally applicable to macroscopic systems and are not constrained by the box size used in simulations or the small time scales of atomic motions. Within the framework of a semi-empirical force field or coarse-grained model, a wide range of liquid-state theories are available (Figure \ref{fig:tree}). These theoretical techniques enable efficient analysis of the properties and phase behavior of molecular systems without the need to explicitly sample the microstates.  Moreover, they provide insights into the intrinsic relationships between various thermodynamic properties and microscopic interactions, which can be challenging to extract from molecular simulation. For these reasons, liquid-state theories can serve as a powerful alternative to simulation methods for predicting the properties of molecular systems, particularly in dealing with large or complex systems where exhaustive simulation may be computationally prohibitive.

The shortcomings and limitations of liquid-state theories are also well known. Liquid-state theories often resort to perturbation expansions or mean-field approximation to describe thermodynamic non-ideality arising from correlation effects. These approximations may introduce unpredictable errors in comparison with exact results. While the analytical methods offer advantages in terms of computational efficiency, their numerical implementation can still be demanding, particularly when dealing with highly inhomogeneous and/or asymmetric systems entailing both large and small molecules. To some extent, finding a delicate balance between accuracy and computational efficiency is akin to approaches taken in applying the \ac{KS-DFT} for electronic structure calculations. While there have been substantial endeavors to enhance the accuracy and computational efficiency of \ac{KS-DFT} by incorporating \ac{ML} methods,\cite{RN8391}  comparable efforts are only in their nascent stages in the realm of liquid-state theories. In this article, we highlight some of the recent advancements in this field by using integral-equation theories and classical density functional theory (cDFT) of simple fluids as illustrative examples. We outline a few future research directions aimed at improving the accuracy and computational efficiency of liquid-state calculations through  incorporating recent advancements in machine intelligence, such as 
{{surrogate models}}, 
uncertainty quantification, and dimension reduction techniques. 

\section{Integral-equation theories}
Integral-equation theories provide formal mathematical relationships among various correlation functions underlying the non-ideal behavior of thermodynamic properties of a many-body system.\cite{RN8359}  One prime example is the \ac{OZ} equation, which describes the spatial correlations of local particle densities at different positions. For a uniform system of spherical particles with average density $\rho$, the \ac{OZ} equation can be written as
\begin{align}  \label{equ:OZE}
h(r) = c(r) + \rho \int d \br' c(|\br - \br'|) h(r') 
\end{align}
where $h(r)$ and $c(r)$ are the total and direct correlation functions, respectively. The total correlation function $h(r)$ is a dimensionless quantity depending on the radial distance $r$ from the position of any tagged particle in the system. It depicts the (normalized) local density of other particles relative to the average density. Intuitively, Eq.\eqref{equ:OZE} suggests that $h(r)$ comprises a direct contribution $c(r)$ accounting for the correlated particle densities at two positions separated by distance $r$, and indirect contributions due to correlation with particle densities at all other positions $\br'$ in space. The 3-dimensional Fourier transforms of the total and direct correlation functions, $\hat{h}(\bk)$ and $\hat{c}(\bk)$, are related to the static structure factor $\hat{\mathcal{S}}(\mathbf k)$
\begin{align} \label{equ:hgs}
\hat{\mathcal{S}}(\bk)=1+\rho \hat{h}(\bk) =\frac{1}{1-\rho \hat{c}(\bk)}
\end{align}
The static structure factor is directly proportional to the scattering intensity measured in neutron and X-ray scattering experiments, rendering it indispensable for analyzing the experimental data. Assuming a pairwise additive potential $u(r)$ for particle-particle interactions, one can readily predict all thermodynamic properties of the one-component system from the total and direct correlation functions using standard statistical-mechanical equations.  

Integral-equation theories are theoretically rigorous and broadly applicable to mixtures as well as complex fluids.\cite{gray2011theory} Because the correlation functions enable the prediction of both microscopic structure and thermodynamic properties without the need to explicitly consider microstates, integral-equation theories have the unique capability to describe the properties of macroscopic systems from a molecular perspective. However, one caveat is that, in practical applications, additional information is needed for describing the intermolecular potential and its connection with the correlation functions. For example, the \ac{OZ} equation for one-component systems of spherical particles discussed above requires an additional relation linking the total and direct correlation functions to the pair potential $u(r)$
\begin{align}  \label{equ:Closure} 
\ln [h(r)+1] = -\beta u(r) + h(r) -c(r) + b(r)
\end{align}
where $\beta=1/(k_BT)$ with $k_B$ being the Boltzmann constant and $T$ absolute temperature, and $b(r)$ denotes the bridge function. The latter is defined by the high-order terms in the diagrammatic expansion of the total correlation function when it is expressed in terms of the ensemble average or, equivalently, the functional expansion of the free energy with respect to the local particle density.\cite{RN8359} Without $b(r)$, Eq.\eqref{equ:Closure} is known as the hypernetted chain (HNC) approximation. 

In integral-equation theories, the additional equation relating the total and/or direct correlation functions to the intermolecular potential, as given by Eq.\eqref{equ:Closure} for one-component systems, is commonly known as the closure. Whereas a formal relation among these functions can be derived for any molecular system, an exact closure involves a bridge function that is not analytically tractable even for simple fluids. The practical application of the \ac{OZ} equation mostly relies on approximate closures derived through heuristic analysis or the universality \textit{ansatz} for the bridge function. Over the years, highly accurate closures have been developed for various systems, applicable not only to simple fluids but also to complex systems such as aqueous solutions of electrolytes, molten salts, and polymeric materials.\cite{PELLICANE2020112665,10.1063/5.0070869,doi:10.1021/cr5000283} Importantly, analytical results derived from the Percus-Yevick (PY) equation and the mean-spherical approximation (MSA) play a pivotal role in the development of modern liquid-state theories for both bulk and inhomogeneous fluids. 

While the importance of integral-equation theories cannot be overstated, their widespread application is impeded by challenges in numerically solving the correlation functions and the inherent inaccuracies of approximate closures. These limitations may result in thermodynamic inconsistencies and nonphysical predictions, hindering their extensive utilization. These issues have been recognized for decades yet with only limited progress. It was only recently demonstrated that both numerical and theoretical challenges can be mitigated by leveraging on machine intelligence. 

\subsection{\ac{OZ} surrogates}
Statistical and machine learning techniques prove invaluable in overcoming computational bottlenecks associated with solving high-dimensional nonlinear equations. We anticipate that similar methodologies can be employed to alleviate numerical difficulties in solving the \ac{OZ} equation. 

\begin{figure*}[ht]
    \centering
    \includegraphics[scale=0.85]{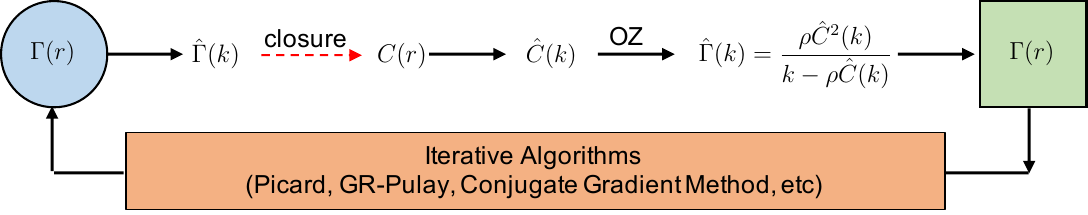}
    \caption{Solving the Ornstein-Zernike (OZ) equation with iterative methods is computationally demanding and frequently beset by convergence. Here, $\Gamma(r) \equiv r[h(r) -c(r)]$, $C(r)\equiv rc(r)$, $\hat{\Gamma}(k)$ and $\hat{C}(k)$ represent the Fourier transforms of $\Gamma(r)$ and $C(r)$.}
    \label{fig:SCIteration}
\end{figure*}

As shown schematically in Figure \ref{fig:SCIteration}, conventional methods for solving the \ac{OZ} equation start with an initial guess for the indirect correlation function $\Gamma(r) \equiv r[h(r) -c(r)]$, followed by the calculation of the direct correlation function from a closure. Next, the \ac{OZ} equation is used to generate a new estimate for $\Gamma(r)$ through the Fourier transform, and the  procedure iterates with an appropriate numerical scheme until the solution reaches convergence.  While the computational cost is dependent on the complexity of the system under consideration and numerical details, one major obstacle in practical applications is that the \ac{OZ} equation may not have a numerical solution, or the iterative process fails to converge.    

Recently, Carvalho and Braga demonstrated that, for the Lennard-Jones fluids, the \ac{OZ} equation  can be alternatively solved by using Physics Informed Neural Networks (PINN).\cite{CARVALHO2022120504} Regrettably, the authors reported that the \ac{ML} model took longer computational time than the direct iterative methods for solving the \ac{OZ} equation with the PY and HNC closures. Nevertheless, PINN has advantages when considering additional physical information such as the asymptotic behavior of the correlation functions near the critical point. Besides, PINN can be used to retrieve correlation functions from experimental data for the static structure factor. More efficient \ac{ML} techniques have been proposed to solve the \ac{OZ} equation by combining physics-informed neural operator networks with Fourier feature embedding and a self-adaptive weighting method.\cite{chen2023physics} The surrogate models for solving the \ac{OZ} equation can be similarly applied to other integral-equation theories including those for complex molecular systems. By establishing \ac{ML} correlations between input and output functions directly, the integral-equation theories could facilitate rapid predictions of the equilibrium structure, solvent-mediated interactions and phase behavior of polymeric materials and biochemical systems.\cite{HALL201038,doi:10.1021/acs.jpcb.2c03384,thermo3030023} 

Neural network approaches like PINN integrate both the fit of simulation runs  and the governing equations for learning the latent function. However, they do not directly provide uncertainty quantification as the parameters are optimized using a loss function. 
Another strategy of building surrogate models\cite{santner2003design} is to learn the mapping from inputs, such as the pair potential and average density in the OZ equation, through Bayesian inference.  Bayesian models, such as Gaussian process (GP) emulators,\cite{deringer2021gaussian,gramacy2020surrogates} 
have been widely used  for predicting computationally expensive  simulations in  different disciplines.  
One advantage of Bayesian modeling is that the uncertainty of the predictions can be quantified from the predictive intervals, either through posterior sampling or approximation from the  posterior mode estimation, whereas  a neural network approach may not  directly provide assessment of the uncertainty in predictions. Furthermore, a GP emulator has a closed-form predictive distribution and it does not have many tuning parameters as neural networks. These properties    make the GP emulator less data-demanding and easier to train. However, the computational complexity of a GP emulator  increases cubically with the number of samples in the training data set. When dealing with a large number of training samples, techniques such as the inducing point approach\cite{snelson2005sparse} are often used to approximate the likelihood function for training the GP emulator with a reduced computational cost.


\subsection{\ac{ML} closure}
\begin{figure*}[ht]
    \centering
    \includegraphics[scale=0.8]{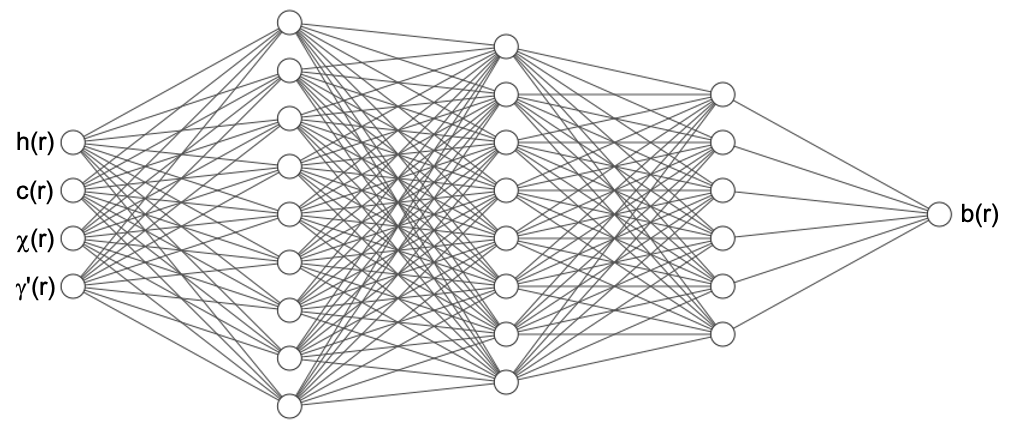}
    \caption{Schematic of a multilayer perceptrons neuron network (MPNN) for inferring $b(r)$ from $h(r), c(r), \chi(r)$ and $\gamma'(r)$. Reproduced from Goodall and Lee\cite{RN387}.}
    \label{fig:MPNN}
\end{figure*}

The total and direct correction functions of a molecular system can be extracted from molecular simulation.\cite{doi:10.1080/00268976.2016.1143567} Subsequently, the simulation data can be utilized to calculate and train the bridge function based on the pair potential and the exact closure relation, \eg,  Eq.\eqref{equ:Closure} for one-component systems of spherical particles. Goodall and Lee carried out \ac{MD} simulation for thirteen one-component systems of spherical particles over a broad range of thermodynamic conditions and inferred the bridge functions using a class of neural networks called multi-layer perceptrons (MLP).\cite{RN387} As illustrated schematically in Figure \ref{fig:MPNN}, the bridge function can be regressed in terms of a set of structure features including not only the total and direct correlation functions $h(r)$ and $c(r)$ but also the pair fluctuation function $\chi(r)$ and the gradient of the indirect correlation function $\gamma'(r)\equiv h'(r)-c'(r)$. The \ac{ML} closure was tested by applying it to solve the so-called inverse problem of statistical mechanics, \ie, extracting the pair potential from the simulation data for the static structure factor. These authors found that the \ac{ML} approach yields better predictions of the pair potential than conventional closures such as PY and HNC. It was suggested that even more robust and reliable predictive capability may be achieved through  Bayesian neural networks. 

The \ac{ML} closures derived by Goodall and Lee {{corroborate}} the universality \textit{ansatz} for the bridge function, \ie, the bridge function of a simple fluid is mainly determined by the short-range repulsion. However, it was not clear whether the learned bridge functions could be represented by effective hard-sphere systems as commonly assumed in the literature. Further development of \ac{ML} closures is needed to correlate the bridge function directly with the potential energy and for systems exhibiting more complicated intermolecular interactions.  It is probably worth noting that the ML closure approximates a complex functional whereby accurate interpolation in high-dimensional space can be challenging for test inputs that are far away from the training data set. The predictive accuracy of ML closures can be improved by reducing the dimension of the input or utilizing quantified uncertainty from a probabilistic model to control the predictive error, which has been demonstrated for emulating classical density functional theory.\cite{10.1063/5.0121805}

\subsection{Structure-centered modeling}
Typically, classical force fields employ pairwise additive potentials to describe intermolecular interactions. Despite the apparent deviations of the effective pair potential from the interaction between two isolated molecules in a vacuum, classical models can achieve remarkable accuracy in predicting the microscopic structure and thermodynamic properties of condensed-matter systems, given that the model parameters have been carefully calibrated. In this context, integral-equation theories can be used to solve the inverse problem mentioned above. With the ground truth on microscopic structure obtained from \ac{AIMD} or scattering experiments, \ac{ML} methods can be employed to improve not only the closure relations but also the quality of force-field parameters for representing diverse properties of the system under consideration. 

 \begin{figure*}[ht]
    \centering
    \includegraphics[scale=0.6]{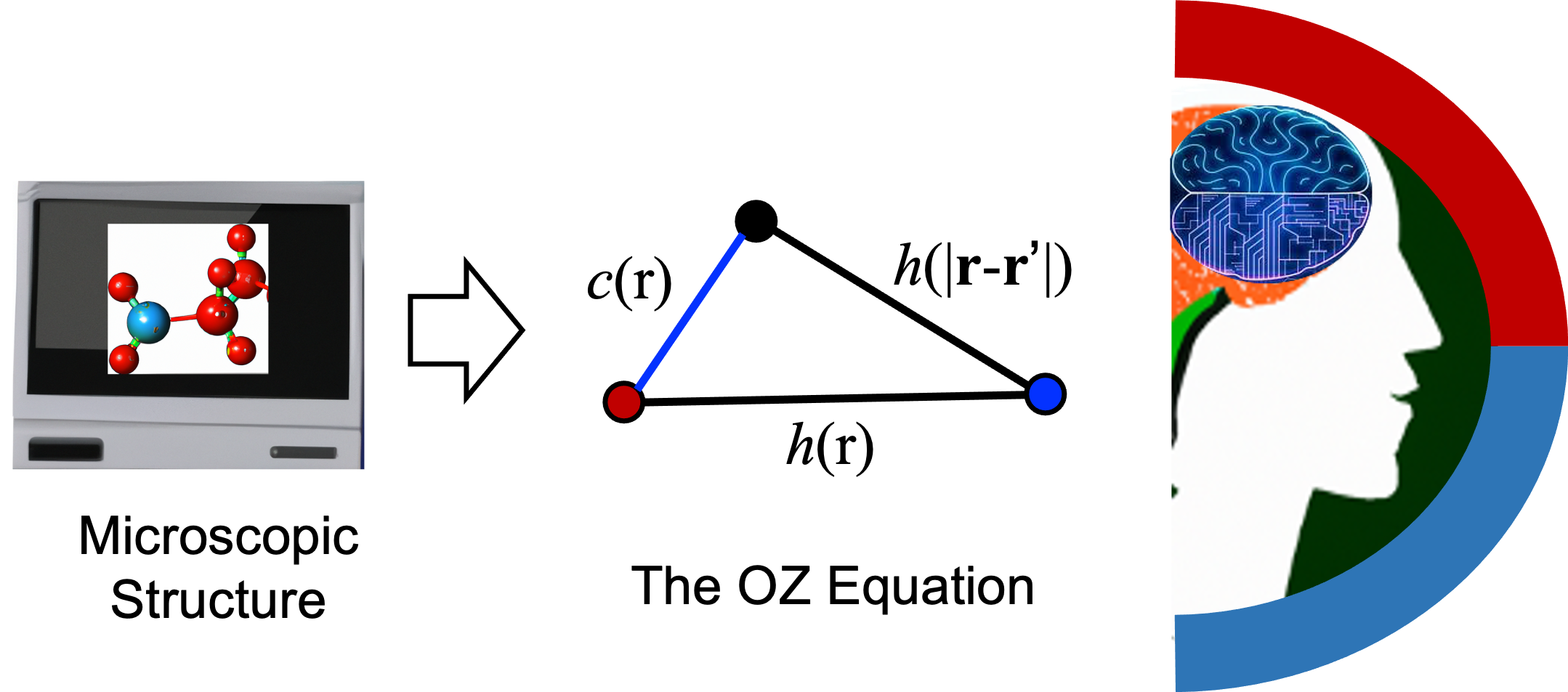}
    \caption{Predicting the properties of condensed-matter systems through pairwise potentials derived from the microscopic structure.}
    \label{fig:EPP}
\end{figure*}

Figure \ref{fig:EPP} shows schematically the ``structure-centered'' approach for predicting the properties of condensed-matter systems. Rather than seeking multibody or \ac{ML} potentials, structure-property relationships can be established by assuming \textit{a priori} pairwise additivity for the effective pair potential. While the state-dependent potential may have its own limitations in describing multi-body interactions, it can be optimized to reproduce microscopic structure as well as experimentally observed thermodynamic properties. {{Importantly, the effective potential eschews problems associated with force field transferability (and scalibility) because the molecular model is constructed for the specific system under consideration.}} Recently, there have been significant advances in using \ac{ML} methods to extract information on the microscopic structure from molecular simulation and scattering experiment, making it easier to calculate the pair distribution functions and the effective pair potentials.\cite{doi:10.1021/acs.jpclett.8b03872} 
Impressively, the so-called point cloud-based deep learning strategy is able to rapidly predict the equilibrium structure of a liquid system from a single molecular configuration.\cite{doi:10.1021/acs.jcim.3c00472}.

When estimating the parameters or functions in these \ac{ML} methods, one often fits the model by minimizing the difference between the modeled output and transformed experimental data at a selected regime or reweighted data based on a loss function. However, the uncertainty of the transformed data  from measurement noise or model discrepancy may not be always homogeneous or independent,  particularly for scattering measurements, which could make some seemingly sensible choices of loss functions, such as the squared error loss, not optimal in estimation. Bayesian  inference, on the other hand, often starts with a probabilistic generative model,\cite{gu2023ab,gu2023probabilistic}  which can improve the loss-minimization estimation through deriving more efficient  estimators based on the probabilistic model. Having a generative model also enables uncertainty estimation to be quantified, and thus enhancing our understanding of implicit model assumptions made in the loss function.



\section{Classical density functional theory}
\label{sec:cdft}
{{While integral-equation theories focus on total and direct correlation functions, \ac{cDFT} aims to predict thermodynamic properties directly from one-body density profiles.\cite{Evans79} 
These methods are formally interconnected through variational calculus. In principle,  all expressions in integral-equation theories can be derived from \ac{cDFT}, but not \emph{vice versa}, because the latter deals with the free energy instead of pair correlation functions. Although integral-equation theories have also been utilized for inhomogeneous systems, they do not rely on the minimization of a free energy and are mostly concerned with the structure of bulk systems. In contrast, the functional dependence of free energy on one-body density profiles and external potentials is an essential component of \ac{cDFT}, making it naturally suitable for studying inhomogeneous systems.\cite{RN8359} }}       

Conceptually, \ac{cDFT} shares the mathematical foundation with \ac{KS-DFT} and they face similar challenges in practical applications. Both methods are formally exact but constrained by the imprecise formulation of the density functional and the numerical complexity in solving the DFT equations.  While \ac{KS-DFT} predicts the ground-state energy of an electronic system by mapping the density into that of ideal fermions, \ac{cDFT} is mainly concerned with the grand potential of a thermodynamic system as a functional of the density profiles of the underlying species. Minimization of the grand potential leads to the Euler-Lagrange equation, which for a one-component system of spherical particles, is given by
\begin{align}
    \rho(\br) = \frac{1}{\Lambda^3}\exp \Big\{ - \beta \fai(\br) + c_1(\br)  \Big\}
    \label{equ:rho_EL}
\end{align}
where $\Lambda$ stands for the de Broglie thermal wavelength, $\fai(\br) \equiv V^{ext}({\br}) -\mu$ is defined as the external potential $V^{ext}({\br})$ minus the chemical potential $\mu$, and $c_1(\br)$ represents the one-body direct correlation function. Eq.\eqref{equ:rho_EL} is analogous to the KS equation, \viz, it predicts the density profile of a non-interacting system under an effective one-body potential. Similar \ac{cDFT} equations can be written for molecular and polymeric systems.\cite{VALIEV2022108338,10.1063/5.0022568,10.1063/5.0057506,PIZIO2023123009,Ginzburg2021,Forsman2017} 

Ottel and coworkers first explored the feasibility of using neural networks to construct free-energy functionals for one-dimensional hard rods and Lennard–Jones fluids.\cite{10.21468/SciPostPhys.6.2.025, 10.1063/1.5135919} With the density functionals expressed in terms of a polynomial or artificial neural networks, they found that the \ac{ML} functionals are able to reproduce the density profiles at conditions beyond the training region. Yatsyshin \etal  introduced a Bayesian  method for predicting the external potential of classical particles from the density profiles or configurations generated through MC simulation.\cite{10.1063/5.0146920,10.1063/5.0071629} The predictions from the \ac{ML} regressions were found in good agreement with the exact results for both the grand potential and density profiles of one-dimensional hard-rod systems. Cats \etal combined the \ac{cDFT} formalism with a stochastic optimization method to regress the non-mean-field component of free-energy functional for the Lennard-Jones fluids.\cite{10.1063/5.0042558} They expressed the non-mean-field free energy in terms of the density profiles in quadratic and cubic forms with the corresponding kernels obtained from fitting with density profiles obtained from grand-canonical MC simulation. The \ac{ML} functional also provides good predictions of the thermodynamic properties and pair correlation functions of bulk fluids.

In the aforementioned publications, the density profiles are calculated through an exact functional or molecular simulation under different external potentials. Subsequently, \ac{ML} methods are employed to regress the excess Helmholtz energy as a functional of the one-body density profile with a predetermined form. More recently, Samm\"uller \etal demonstrated that a deep neutral network can be constructed to regress the functional dependence of the one-body direct correlation function with respect to the density profile.\cite{sammüller2023neural} Subsequently, the functional relation between $c_1(\br)$ and $\rho(\br)$ can be utilized for the calculation of the grand potential by functional line integration 
\begin{align}
   \beta \Omega= \int d\br \rho(\br) \Big\{ \ln [\rho(\br)\Lambda^3] -1 + \beta \fai(\br) \Big\} - \int_0^1 d \lambda \int d\br \rho(\br) c_1(\br; \lambda\rho(\br)).  
    \label{equ:Omega}
\end{align}
Impressively, the neural functional was found outperforming the state-of-the-art \ac{cDFT} methods for inhomogeneous systems of hard-sphere fluids. 

 \begin{figure*}[ht]
    \centering
    \includegraphics[scale=0.6]{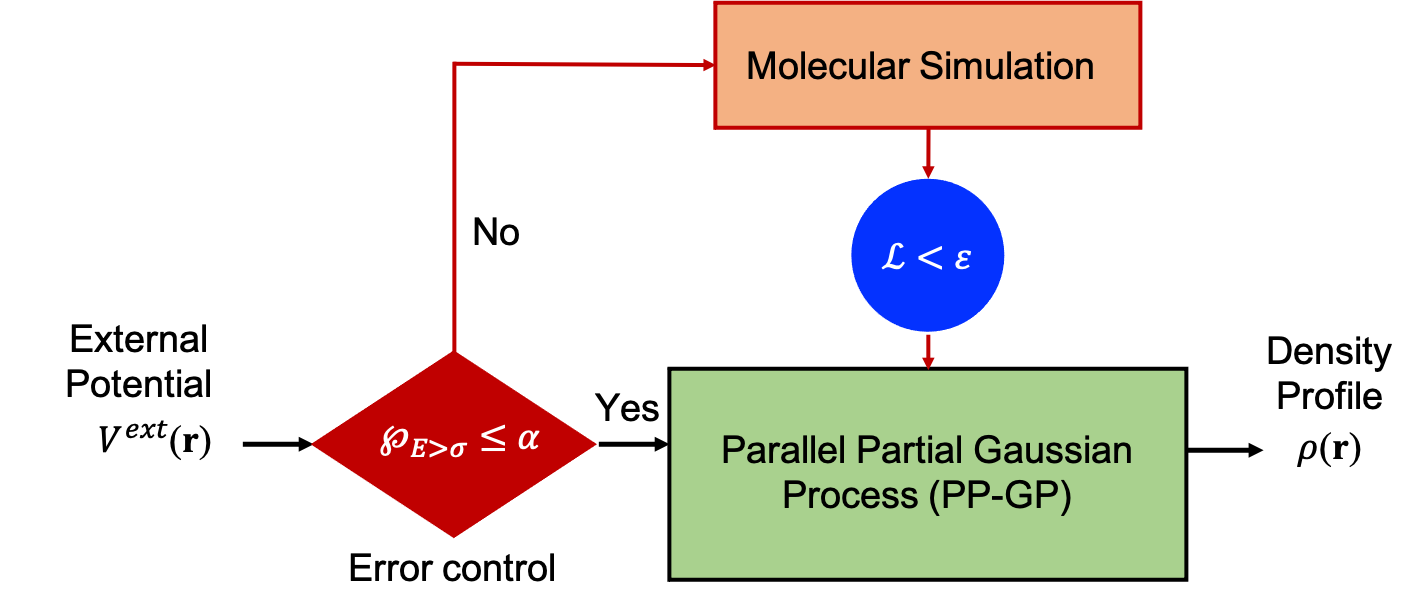}
    \caption{Schematic of active functional learning with error control. Here, the functional learning approach is applied to regress the dependence of the one-body density profile on the external potentials, $ \powerset_{E > \delta} \leq \alpha$ stands for the probability of the predictive error larger than $\delta$ is less than $\alpha$, and $\mathcal{L}<\varepsilon$ means the loss function is less than $\varepsilon$.}
    \label{fig:AlecDFT}
\end{figure*}

Whereas the \ac{ML} techniques discussed above were tested only with simple systems, we expect that similar improvements can be achieved for all cDFT calculations by formulating suitable neutral functionals. In the application of cDFT to molecular systems, the numerical procedure will be necessarily more complicated and requires more advanced \ac{ML} methods such as active learning techniques and dimensionality reduction. Although cDFT calculations are faster than molecular simulation, minimizing the grand potential or solving the Euler-Lagrange equation is computationally demanding for inhomogeneous systems containing polymeric species or molecules interacting through fine-detailed potentials. With an expression for the grand potential, which can be obtained either from theoretical derivation or machine learning, surrogate \ac{ML} models can be established to predict density profiles with the external potential as the input. For example, artificial neural network was used as a surrogate for the iterative solution of the Euler-Lagrange equation for liquid-crystal systems under confinement.\cite{PhysRevE.89.053316}. More recently, Fang \etal proposed 
an active learning framework that utilizes a parallel partial Gaussian process (PP-GP) emulator \cite{gu2016parallel} to predict $\rho(\br)$ from external potential $V^{ext}({\br})$,  and demonstrated that the \ac{ML} approach is more computationally scalable than direct cDFT calculations.\cite{10.1063/5.0121805}. As shown schematically in Figure \ref{fig:AlecDFT}, the active learning approach is able to discern the functional inputs that can be reliably predicted based on quantified uncertainty from the emulator, thereby controlling the predictive error in a high-dimensional input space for new applications. 


\section{Summary and Perspectives} 
Liquid-state theories play an instrumental role in predicting the thermodynamic properties and phase behavior of liquid mixtures and solutions important for diverse scientific fields. Whereas approximations are inevitable in conventional practice, their integration with 
\ac{ML} methods holds the potential for more scalable and accurate calculations over a wide range of conditions. In particular, the theoretically-guided functional 
learning methods will eventually offer a valuable alternative to molecular simulations for mapping the high-dimensional potential energies of thermodynamic systems into equilibrium properties of practical concern. 

Machine learning (\ac{ML}) methods such as Gaussian Processes (GP) and Physics-Informed Neural Networks (PINN)  have been widely used for emulating complex functions or functionals, and for solving differential and integral equations. However, similar to other molecular-level simulations, a fundamental bottleneck arises from the high-dimensionality of input space when approximating liquid-state theories. Two promising directions are possible for improving the application of ML methods. First, reducing the dimension of the input space has the potential to  significantly lower the computational cost and improve predictive accuracy. Some physics equations intrinsically depend on a low dimensional space of the entire input space. For example, the one-body direct correlation function $c_1(\br)$ of a hard-sphere system is a short-ranged function of one-body density profile $\rho(\br)$. In this case, we can build a ML surrogate model for $c_1(\br)$ using the local density instead of the entire density profile as the input. The input dimension can be further reduced by using the active subspace (AC) approach,\cite{constantine2014active} which projects the high-dimensional input vector to a few principal axes that maintain the largest deviation from the  output gradients.
 The AC approach is often more accurate to capture the input-output map than unsupervised approaches such as the principal component analysis, as the output variation characterized by the gradients are used for building principal axes. 

In addition to dimension reduction, having reliable uncertainty assessment from ML techniques can be useful to improve the predictive accuracy, particularly for extrapolating the input space. The predictive intervals generated by Bayesian models, such as the parallel partial Gaussian process (PP-GP) emulator \cite{gu2016parallel} and corresponding software package \cite{RJ-2019-011}, can be used to determine whether particle density and grand potential in cDFT calculations can be reliably predicted for a new external potential.  \cite{10.1063/5.0121805} These intervals help control the predictive errors in a probabilistic manner, offering valuable insights into the accuracy of the predictions. 
In ML techniques such as artificial neural networks, predictive uncertainty may not be directly available because parameters are estimated by minimizing a loss function, instead of defining a generative model of the data. Building a probabilistic generative model enables the uncertainty to be quantified. Moreover, the generative model can motivate the derivation of more efficient estimators that improve estimation from loss-minimization.\cite{gu2023ab}  Therefore, dimension reduction and uncertainty quantification from probabilistic generative models are two critical directions to enhance the performance of liquid-state theories for computationally challenging scenarios, such as 3D systems with long-range intermolecular correlations. 

Addressing the computational challenges associated with bridging molecular and macroscopic scale phenomena has been a fundamental pursuit in theoretical research within the realms of chemistry and materials science. Surrogate models and functional learning  methods have proved valuable for both quantum and statistical mechanical calculations and are expected to play an increasingly important role in advancing this endeavor.

\bibliography{PLST}        
\bibstyle{chem-acs}
\end{document}